\documentclass[aps,prd]{revtex4}
\input{epsf}

\newcommand{\be}{\begin{equation}}
\newcommand{\ee}{\end{equation}}
\newcommand{\bea}{\begin{eqnarray}}
\newcommand{\eea}{\end{eqnarray}}
\newcommand{\ba}{\begin{array}}
\newcommand{\ea}{\end{array}}

\begin{document}

\title{Regularizing the r-mode problem for nonbarotropic relativistic stars}

\author{Keith H. Lockitch}
\affiliation{Department of Physics, University of Illinois Urbana-Champaign,
1110 E. Green St., Champaign-Urbana, IL 61801, USA}
\author{Nils Andersson}\email{na@maths.soton.ac.uk}
\affiliation{School of Mathematics, University of
Southampton, Southampton, SO17 1BJ, UK}

\author{Anna L. Watts}\email{anna@milkyway.gsfc.nasa.gov}

\affiliation{Laboratory for High Energy Astrophysics, NASA Goddard
  Space Flight Center, Greenbelt, MD 20771, USA}

\begin{abstract}
We present results for r-modes of relativistic nonbarotropic stars. 
We show that the main differential equation, which is formally singular
at lowest order in the slow-rotation expansion,
can be regularized if one considers the initial value problem rather
than the normal mode problem.  However, a more physically motivated way to
regularize the problem is to include higher order terms.  
This allows us to develop a
practical approach for solving the problem and we provide results
that support earlier conclusions obtained for uniform density stars. 
In particular, we show that there will exist a single r-mode 
for each permissible combination of $l$ and $m$. We discuss 
these results and provide some caveats regarding their usefulness
for estimates of gravitational-radiation reaction timescales. 
The close connection between the seemingly singular
relativistic r-mode problem and issues arising because of 
the presence of corotation points in differentially rotating stars 
is also clarified.  
\end{abstract}

\maketitle

\section{Introduction}

In the last few years the instability associated with the 
r-modes of a rotating neutron star has emerged as a 
plausible source for detectable gravitational 
waves. 
This possibility has inspired a considerable amount of 
work on gravitational-wave driven instabilities in rotating 
stars and our understanding continues to be improved 
as many of the relevant issues are intensely scrutinized
(see \cite{fl1,ak,lindblom,fredlamb,fl2, narev} for detailed reviews 
and important caveats on the subject).
To date, most models for the unstable r-modes are based on
Newtonian calculations and the effect of the instability on
the spin rate of the star is estimated from post-Newtonian
theory. This may seem peculiar given that the instability is a truly 
relativistic phenomenon (its driving mechanism is gravitational
radiation reaction); but a complete relativistic calculation of 
the oscillation modes of a rapidly rotating stellar model 
(including the damping/growth rate due to gravitational-wave
emission) is still outstanding, and the inertial modes of 
relativistic stars (of which the r-modes form a sub-class) have 
actually not been considered at all until very recently. 
In contrast, our understanding of  
rotating Newtonian stars has 
reached a relatively mature level and it is thus not 
surprising that most attempts to understand
the r-mode instability and its potential astrophysical 
relevance have been in the context of Newtonian theory.

Table~\ref{tab1} summarizes the differences between the low frequency
modes of barotropic and nonbarotropic stars, and the  
the ways in which the relativistic inertial mode problem differs from 
the Newtonian problem (primarily because of the dragging of inertial 
frames)~\cite{laf1}.  Following \cite{laf1}, we use the term
barotropic to describe a star for which the true equation of state describing both the  
background star and its perturbations is a prescribed one-parameter 
function $p=p(\epsilon)$.  In a nonbarotropic star the perturbations
and background star obey different equations of state.  The main cause
of nonbarotropy in neutron stars is stratification via entropy or
chemical composition gradients, with the latter being the most 
important for all but very hot (newly born) stars.

\begin{table}[h]
\begin{tabular}{|l|l|l|}
\hline
 & Nonbarotropic stars & Barotropic stars \\
\hline
{\bf Newtonian Theory } & infinite set~\cite{orders} 
of r-modes for each $[l,m]$ 
& a single r-mode for $l=m$\\
& infinite set of g-modes & infinite set of inertial modes \\
\hline
{\bf General Relativity} & infinite set~\cite{orders} 
of r-modes for each $[l,m]$ 
& no pure r-modes \\
& infinite set of g-modes &  infinite set of inertial modes \\
& continuous spectrum? & \\
\hline
\end{tabular}
\caption{A comparison of inertial mode results in Newtonian 
gravity and General Relativity.}
\label{tab1}
\end{table}

First, some notes on terminology. Perturbations of a spherical star can be 
decomposed into two classes depending on how the perturbed velocity
transforms under parity (see \cite{laf1}). Following the standard relativistic terminology we 
will refer to perturbations that transform under parity like the 
scalar spherical harmonic $Y_l^m$ as ``polar'',
while referring to those that transform opposite to $Y_l^m$ as 
``axial''.   This classification applies also to rotating stars, 
with the parity class of a mode being determined by its spherical limit
along a sequence of rotating models~\cite{modeparity}.

Let us now characterise the rotationally restored modes; modes that
have zero-frequency in the non-rotating (spherical) limit.  We refer
to modes that become purely axial in the spherical limit as r-modes,
while modes that limit to a mixed polar/axial parity state are called
inertial modes.  Rotation breaks the degeneracy of these modes and
gives them a finite frequency that is proportional to the star's
angular velocity $\Omega$.  We will also consider the g-modes, modes
that are restored by the buoyancy associated with stratification.  

A rotating Newtonian barotrope possesses a vestigial set of $l=m$
r-modes and a set of inertial modes, but no g-modes.  A non-barotropic
star, by contrast, possesses r-modes and g-modes.  It is however worth
noting that the low-frequency modes of a 
rapidly spinning neutron star may be similar to those 
of a barotropic model even though 
one would expect a realistic model to be stratified. If the Coriolis force
dominates the buoyancy force, one would expect the low-frequency 
mode spectrum to be made up of inertial modes~\cite{yl}. Given that the g-modes of a ``typical''
neutron star model have frequencies below a few hundred Hz, it seems plausible 
that the low-frequency modes of millisecond pulsars will, in fact, be
inertial modes.

In Refs.~\cite{laf1} and \cite{lfa}, we discussed the rotationally restored
(inertial) modes of a slowly rotating relativistic star in some detail. 
One of the main results of this work was that these modes have a fundamentally
different character in barotropic versus nonbarotropic stellar
models.  Firstly, the vestigial set of $l=m$ r-modes that one finds in a Newtonian 
barotrope do not exist in a relativistic barotropic star
\cite{laf1}.  One is left with only inertial modes. This is
particularly important because it is the $l=m=2$ r-mode that is  
most likely to dominate the gravitational-wave driven
instability. Secondly, it is possible to find r-modes of a
relativistic nonbarotropic  
star at order $\Omega$ in a slow-rotation expansion.  In
the Newtonian case the r-modes are degenerate at 
order $\Omega$; one must undertake an order $\Omega^2$ calculation to find 
the eigenfunctions.  This degeneracy is partially split at
order $\Omega$ in the relativistic case, allowing one to compute
the leading part of the mode eigenfunction.  In the rest of this paper
we will focus
our attention on the relativistic problem  
for nonbarotropic stars, since the barotropic case
is comparatively well understood and has been discussed elsewhere
\cite{lfa}.  

To lowest order in the slow rotation approximation, the r-modes of 
nonbarotropic stars are governed by an ordinary differential equation 
first derived by Kojima \cite{koj}.  Consider the differential
equation  (for a complete derivation, see \cite{laf1})
\begin{equation}
(\alpha -\tilde{\omega}) \left\{  e^{\nu-\lambda} {d \over dr} \left[
e^{-\nu-\lambda} { dh \over dr} \right] - \left[ { l(l+1) \over r^2} - 
{4M(r) \over r^3} + 8\pi(p+\epsilon) \right] h  \right\} + 
16 \pi(p+\epsilon) \alpha h = 0
\end{equation}
which determines the axial metric perturbations for 
a ``pure'' relativistic r-mode ($h$ is directly related to 
$\delta g_{t \theta}$).
In the equation $\nu$ and $\lambda$ are coefficients of the 
unperturbed metric, and $\tilde{\omega}$ is defined in terms of the 
relativistic frame-dragging $\omega$ as
\begin{equation}
\tilde{\omega} = { \Omega - \omega(r) \over \Omega}
\end{equation} 
where $\Omega$ is the (uniform) rotation rate of the star.
Furthermore, we have assumed that (in the inertial 
frame) the mode depends on time as
$\exp(i\sigma t)$ and then introduced a convenient
eigenvalue $\alpha$ as
\begin{equation}
\sigma = -m\Omega\left[ 1 - {2\alpha\over l(l+1)} \right] \ .
\end{equation}
Using the fact that \cite{laf1}
\begin{equation}
{d \over dr} ( \nu + \lambda) = 4\pi r e^{2\lambda} (p+\epsilon)
\end{equation}
we have
\begin{equation}
(\alpha -\tilde{\omega}) \left\{ {d^2h \over dr^2} - 4\pi r e^{2\lambda} 
(p+\epsilon) {dh \over dr} -  \left[ { l(l+1) \over r^2} - 
{4M \over r^3} + 8\pi(p+\epsilon) \right] e^{2\lambda} h  \right\} + 
16 \pi(p+\epsilon) \alpha e^{2\lambda} h = 0
\label{kojima}\end{equation}

The above equation was first derived by Kojima \cite{koj}, and our 
previous analysis \cite{laf1} shows that it can be used to determine 
r-modes of a nonbarotropic relativistic star. 
In order for the 
solution to satisfy the required regularity conditions  
at both the centre and at infinity, 
the eigenvalue $\alpha$ must be such that $\alpha -\tilde{\omega}$ vanishes
at some point in the spacetime \cite{laf1}. 
As long as  $\alpha - \tilde{\omega} \neq 0$
inside the star, the problem is regular and one can readily 
solve it numerically. In our previous study we solved the  
problem for uniform density stars and found that the
required eigenvalues were always such that the problem was 
non-singular~\cite{oops}.
There is of course no guarantee that the problem will remain regular
for more realistic equations of state. Indeed, 
recent work by Kokkotas and Ruoff \cite{kr,kr2} and Yoshida \cite{yosh}
(see also \cite{yf}) extends the analysis to more realistic equations of state, 
such as polytropes, and shows that the desired eigenvalue is then not 
generally such that $\alpha > \tilde{\omega}_s \equiv \tilde{\omega}(R)$. 
In other words, one is (at
least for some stellar parameters) forced to 
consider a singular eigenvalue problem.  Our analysis is intended to extend our results \cite{laf1} for 
uniform density stars to more realistic equations of state. 

The existence of a singular eigenfunction problem has two
consequences.  Firstly, it has been argued that the singularity in
Kojima's equation gives rise to a continuous spectrum of axial
perturbations \cite{koj,bk}. One purpose of this paper
is to explore the nature of the continuous spectrum and demonstrate
that it is an artifact of the slow rotation approximation that may not
be present in physical stars. It has also been suggested \cite{kr,kr2,yosh,yf}  
that the r-modes may not exist in stars for which Kojima's equation is singular. 
Our aim in this paper is to argue that regular r-mode solutions will indeed
exist in such stars.

We will draw in part on studies of the oscillations of differentially
rotating Newtonian stars, which also give rise to singular normal mode
equations.   Although the normal mode solutions are singular, the
perturbation that one obtains by solving the initial value problem is
non-singular \cite{w1,w2}.  The singularity is an artifact of assuming a normal
mode time dependence.   That consideration of the initial value
problem regularizes the relativistic slow rotation 
inertial mode equations in the Cowling approximation has
been demonstrated by Kojima and Hosonuma \cite{khp}. We will
show that the same is true for nonbarotropic stars 
when one includes the metric perturbations, despite the fact
that the character of the singular normal mode solutions is different
to that of the solutions found in the Cowling approximation.  

It is also plausible that the 
singular character of Kojima's equation represents a breakdown in the
slow-rotation approximation.  We will argue that one may regularize 
the singular normal mode equation in a physically well-motivated way by including 
higher order terms in the approximation. This issue, which 
develops further the work of Kojima and Hosonuma
\cite{kh}, was addressed in
some detail in an earlier preprint by two of us
\cite{pre}.  This paper is a slightly revised version of that
original preprint. In particular, we incorporate recent improvements in our understanding of
the nature of continuous spectra obtained from studies of
differential rotation~\cite{w1,w2}.

Other methods of regularization that have been discussed
in the literature include the effect of gravitational radiation
reaction, and coupling to higher order multipoles.  For the
nonbarotropic problem, gravitational radiation reaction alone is not
sufficient to regularize the singular solutions found at lowest order
in the slow rotation expansion \cite{yf,kr2}. Coupling to higher order
multipoles, however, does regularize the nonbarotropic problem both
in the Cowling approximation \cite{yl2} and when one includes the
metric perturbations \cite{kh}.  A brief note on the occurrence of
continuous spectra and regularization in the barotropic problem is
included in the Appendix.

\section{Singular eigenfunctions}
\label{singe}

Let us consider Eqn. (\ref{kojima}) in the case when $\alpha$ is such 
that we have 
\begin{equation}
\alpha - \tilde{\omega}(r_0) = 0
\end{equation}
for $r_0<R$, 
i.e. when the problem is singular at some point $r_0$ in the stellar fluid. 
Suppose we use a power series expansion to analyze the behaviour 
of the solutions to (\ref{kojima}) in the vicinity of the singular point.
Expanding in $x=r-r_0$, and assuming that all quantities that describe the
unperturbed star are smooth, we use
\begin{eqnarray}
\tilde{\omega} &\approx& \tilde{\omega}(r_0) + 
\tilde{\omega}_1 x + \tilde{\omega}_2 x^2 ... \quad 
\mbox{ where } \alpha - \tilde{\omega}_0(r_0) = 0 \\
p & \approx & p_0 + p_1 x + ... \qquad
\epsilon  \approx  \epsilon_0 + \epsilon_1 x + ... \\
e^\lambda &\approx & e^{\lambda_0} \left[ 1 + \lambda_1 x + ...  \right] \qquad
M  \approx M_0 + M_1 x + ... 
\end{eqnarray}
Next we introduce the Frobenius Ansatz
\begin{equation}
h = \sum_{n=0}^\infty a_n x^{n+\beta}
\end{equation}
in (\ref{kojima}) and find that we must have either $\beta =0$ or $\beta=1$. 
This problem 
thus belongs to the class where the difference between the two values for 
$\beta$ is an integer and we would not expect the two power series
solutions to be independent. Indeed, further scrutiny
of the problem reveals that we can only find one regular power series
solution to our problem. This leads to an approximate solution
\begin{equation}
h^{\rm reg} \approx a_0 x \left[ 1 + a_1 x \right]
\label{regsol}\end{equation}
where
\begin{equation}
a_1 =  2\pi (p_0 + \epsilon_0) e^{2\lambda_0}
\left( r_0 + {4\alpha \over \tilde{\omega}_1} \right) 
\end{equation}

In order to arrive at a second, linearly independent, solution we 
resort to the standard method of variation of parameters. 
Given one solution $h_1(r)$ to (\ref{kojima}), 
a second solution can be obtained as 
\begin{equation}
h_2 = f(r) h_1
\end{equation}
Introducing this combination in (\ref{kojima}) it is straightforward
to show that we must have
\begin{equation}
f^{\prime\prime} h_1 = \left[ 4\pi r e^{2\lambda} (p+\epsilon) h_1 - 2 h_1^\prime \right] f^\prime
\end{equation}
(where a prime denotes a derivative with respect to $r$). In other words
\begin{equation}
{ f^{\prime\prime} \over f^\prime} = 4\pi r e^{2\lambda} (p+\epsilon) - { 2 h_1^\prime \over h_1} 
\end{equation}
which integrates to
\begin{equation}
f^\prime = { D \over h_1^2} 
\exp \left[ \int 4\pi r e^{2\lambda} (p+\epsilon) dr  \right]
\label{fprime}\end{equation}

Unfortunately, in our case we only know $h_1$ in the vicinity 
of the point $r_0$. In order to proceed we therefore expand 
(\ref{fprime}) in terms of $x$ and then use $h_1 = h^{\rm reg}$.
Then we need
\begin{equation}
\exp \left[ \int 4\pi r e^{2\lambda} (p+\epsilon) dr \right] 
\approx 1 + E_1x
\end{equation} 
where $E_1 = 4\pi r_0 e^{2\lambda_0} (p_0+\epsilon_0)$
and
\begin{equation}
\left( {a_0\over h^{\rm reg} } \right)^2 \approx { 1\over x^2} - 
{2a_1 \over x} 
\end{equation}
Putting the various pieces together we have
\begin{equation}
{ df \over dx} \approx C \left\{ { 1\over x^2} + {E_1-2a_1 \over x} 
\right\}
\end{equation}
with $C$ an arbitrary normalisation constant. Integration then yields
(recalling that $x$ can take on both positive and negative values) 
\begin{equation}
f \approx -C \left\{ { 1\over x}  + (2a_1-E_1) \ln |x| \right\} 
\end{equation}
At the end of the day, we have arrived at a second solution to 
our problem (we discuss the consequence of taking $\ln|x|$ rather than
$\ln x$ below). Near the point $r_0$ this solution can be written
\begin{equation}
h^{\rm sing} \approx b_0 \left\{ 1 + b_c x \ln |x| + a_1 x \right\}
\label{singsol}\end{equation}
where
\begin{equation}
b_c = 2a_1-E_1 = {16\pi \alpha (p_0+\epsilon_0) e^{2\lambda_0}
\over\tilde{\omega}_1 }
\end{equation}
(note that we need to keep the last term in (\ref{singsol}) 
to work at an order that allows us to distinguish the leading 
order term of (\ref{regsol})
from the corresponding term in the singular solution).
From this expression it is clear that, while the function
$h^{\rm sing} $ is regular at $r=r_0$ its derivative is 
singular at this point. 

In addition to this, one can show that it is not possible to find an overall 
solution to the problem (that satisfies the required boundary conditions 
at the centre and surface of the star) if one assumes that 
$h\propto h^{\rm reg}$ in the vicinity of $r_0$. 
Given this result we would seem to have two options: 
One option is to conclude that we must have a singular
metric/velocity perturbation, and since this would be unphysical we must 
rule out the associated solution. If we take the implications of this to 
the extreme, it could imply that no relativistic r-modes 
can exist for certain stellar parameters \cite{kr,yosh}. 
However, this conclusion is likely too extreme. It would be surprising
if a small change in, say, the compactness of the star (the stiffness
of the equation of state) could lead to
such a drastic change in the star's physics (the disappearance of
its r-modes).  An alternative (and perhaps more reasonable) option
is to assume that the appearance of a singular eigenfunction signals 
a breakdown in our mathematical description of the problem rather 
than a radical change in the physics.  Later in this paper we will show that the
problem arises because of a breakdown in the slow rotation
approximation.  However, even in the slow rotation approximation, the perturbation
is in fact completely regular; the presence of the singularity in
Kojima's equation is simply a consequence of the assumption of normal
mode time dependence.  

The normal mode equations for differentially rotating
Newtonian stars exhibit mathematically identical singular behaviour
for frequencies that lie within what we call the co-rotation band
\cite{w1,w2}. The eigenfunctions associated with this frequency band
have singular derivatives that possess in general both a logarithmic
singularity and a finite step in the first derivative at the singular
point (see equations (45) - (49) of \cite{w1} and the accompanying 
discussion).  
The additional degree of freedom associated with the
finite step in the derivative permits the existence of a continuous
spectrum of solutions within this frequency band.  At certain frequencies, 
the finite step in the first
derivative vanishes; these frequencies are  referred to as zero-step
solutions and they possess a special character (see below, and the
discussion at the end of Section 6.2 of \cite{w1}).  

The situation for Kojima's equation is identical:  in
general the singular eigenfunctions possess both a logarithmic
singularity and a finite step in the first derivative, leading to a
continuous spectrum of singular solutions.  Just as in the
differential rotation problem, there are certain frequencies for which
the
finite step in the first derivative vanishes.  It can be shown that taking 
the logarithm of $|x|$ in the series
  expansions in Eq.~(20) and demanding continuity of the
  function at the singular point is equivalent to the matching procedure
  used in Section 6.2 of \cite{w1} to pick out the zero-step solutions 
from within
  the continuous spectrum of the differential rotation problem. Thus
  by using $\ln|x|$ rather than $\ln x$ in the
analysis above we are picking out the zero-step solutions from the
continuous spectrum.  

The perturbation is however determined by solution of the
initial value problem rather than the normal mode problem.  Analysis
of the initial value problem for differentially rotating systems
has shown that the perturbation associated with the
continuous spectrum is not singular \cite{w2}.  By  conducting a
similar analysis of
the time-dependent form of Kojima's equation, we have confirmed that the same
is true for the relativistic r-modes. The singular solutions
associated with the continuous spectrum are therefore physically relevant, and
cannot be discounted.

With this in mind, let us review the key characteristics of the
differential rotation continuous spectrum and ask whether similar
characteristics are manifested in the
relativistic problem.  Firstly, the continuous spectrum was found to possess a
position-dependent frequency component; such behaviour has been
observed in numerical time evolutions of the relativistic problem
\cite{kr}. Secondly, there were fixed frequency contributions from the endpoint
frequencies of the continuous spectrum.  Ruoff and Kokkotas \cite{kr}
found such contributions in their simulations, but attributed them to the 
behaviour of the energy density at the surface of the star.  We believe that they
may instead be a hallmark of the continuous spectrum.  The third
characteristic of the continuous spectrum was a power law decay with
time. In \cite{kr} there are two indications of this
type of behaviour.  The amplitudes of the endpoint frequencies
were observed to die away as a power law.  In addition, the authors
noted that there appeared to be no contribution from the continuous
spectrum at late times, suggesting again that it had died away.

Consideration of the initial value problem for differential rotation also
indicated a special role for the zero-step solutions \cite{w2}.  Again, the
 perturbations were found to be non-singular.  For appropriate
initial data the zero-step solutions were found to behave in much the
same way as regular modes outside the co-rotation band, giving rise to
a clear peak in the power spectrum at a fixed frequency and standing
out from the rest of the continuous spectrum.  The zero-step solutions
behaved as modes within the continuous spectrum.  Analysis of the
time-dependent form of Kojima's equation indicates that the same will
be true for the zero-step solutions to the
relativistic problem.  These solutions are therefore relevant.  This
contradicts statements in earlier works \cite{kr,kr2,yosh,yf} that
considered only the normal mode problem.  The authors of these studies
discounted these zero-step solutions within the continuous spectrum as
being unphysical, and concluded that if r-modes entered the continuous spectrum  they
ceased to exist.  In fact they do continue to exist as physically
meaningful zero-step
solutions, and should appear in time evolutions.  For polytropic
background models, Ruoff and Kokkotas \cite{kr} observe no contribution at the
expected zero-step frequency when they initialise their simulations using
arbitrary initial data.  It would interesting to see whether these
modes could be excited using initial data more closely matched to the
zero-step eigenfunction; the zero-step oscillations observed in \cite{w2}
were excited using initial data closely matched to the eigenfunction
rather than arbitrary initial data.  For more realistic equations of
state, however, the time evolutions of \cite{kr} do show clear peaks at fixed
frequencies within the
continuous spectrum.  This suggests the presence of zero-step
solutions.  

Before moving on we should make one comment on the nature of
the continuous spectrum if one makes the Cowling approximation.  In
the Cowling approximation the continuous spectrum eigenfunctions for the velocity
perturbations are delta functions
\cite{khp}. Contrast this to the situation outlined above, where the
velocity perturbations are proportional to the
derivative of the metric perturbation, with a logarithmic
singularity and (in general) a finite step at the singular point. The non-singular
perturbation in the Cowling approximation (found by considering the
initial value problem) exhibits a position dependent frequency
component but no power law time dependence, no endpoint
frequency contributions, and no zero-step solutions.  The nature of
the problem is changed dramatically by working in the Cowling
approximation.  

We have argued above how the singular solutions of Kojima's equation
give rise to non-singular perturbations when one considers
the initial value problem.  However, the main cause of confusion is a
breakdown in the slow-rotation approximation. After all, Eq. (\ref{kojima}) should
really be written 
\begin{equation}
(\alpha -\tilde{\omega}) \left\{ {d^2h \over dr^2} - 
4\pi r e^{2\lambda} (p+\epsilon) {dh \over dr} -  \left[ { l(l+1) \over r^2} - 
{4M \over r^3} + 8\pi(p+\epsilon) \right] e^{2\lambda} h  \right\} + 
16 \pi(p+\epsilon) \alpha e^{2\lambda} h =  0 + O(\Omega^2)
\end{equation}
From this we can immediately see that it is 
inconsistent to use
the slow-rotation expansion 
when $\alpha -\tilde{\omega} \sim O(\Omega^2)$ or smaller.
For the problem at hand this means that the assumptions used in 
the derivation of Eq. (\ref{kojima}) are not consistent in the vicinity 
of $r_0$. Near this point we cannot discard the higher order 
terms while retaining the term proportional to $\alpha -\tilde{\omega}$
since the latter   becomes arbitrarily small. 

At first sight this may seem quite puzzling but 
similar situations are, in fact, common in problems involving 
fluid flows. In such problems, the singularity is usually regularized 
by introducing additional pieces of physics in a
``boundary layer'' near the point $r_0$. A typical example of this, that 
has already been discussed in the context of the r-mode 
instability, is provided by the existence of a viscous boundary
layer at the core-crust interface in a relatively cold neutron star
(see \cite{ak} for an extensive discussion). 
In that case the non-viscous Euler equations adequately 
describe the r-mode fluid motion well away from the 
crust boundary, while the viscous terms are crucial for an 
analysis of the region immediately below the crust. 
In our view, the relativistic r-mode problem leads to a similar 
situation: Well away from the point $r_0$ Eq. (\ref{kojima}) 
leads to an accurate representation of the solution, but if we want
to study the region near $r_0$ we need to include ``higher 
order'' terms in our analysis.  

Unfortunately, this means that it becomes very difficult 
to find a complete solution to the problem. The order $\Omega^2$
perturbation equations for a relativistic star are rather complicated
and have not yet been obtained completely.  But for our present
purposes, we can use partial results in this direction. 
Kojima and Hosonuma \cite{kh} have shown that the next order 
in the slow-rotation expansion 
brings in a fourth order radial derivative of $h$ in Eq. (\ref{kojima}).
Retaining only the principal part of the higher order problem we 
then find that (\ref{kojima}) will be replaced by an equation of form
\begin{equation}
 { \tilde{\omega}^2 g(r)}
\alpha r^2 {d^4h \over dr^4}
+ (\alpha -\tilde{\omega}) \left\{ {d^2h \over dr^2} - 
4\pi r e^{2\lambda} (p+\epsilon) {d h \over dr} -  \left[ { l(l+1)
    \over r^2} -  
{4M \over r^3} + 8\pi(p+\epsilon) \right] e^{2\lambda} h  \right\} + 
16 \pi(p+\epsilon) \alpha e^{2\lambda} h = 0 
\label{higherorder}
\end{equation}
where $g(r)$ contains information about the stellar background ---
in particular the stratification of the star. Most importantly
$g(r_0)\neq 0$ and it is therefore clear that the problem is 
perfectly regular also near the point where $\alpha-\tilde{\omega}=0$.

\section{A suitably simple toy problem}

Our main objective is to argue that one can {\em in principle}
regularize the nonbarotropic r-mode problem. Ideally, we would like to 
find the mode-solutions without actually having to derive the 
relativistic perturbation equations to higher orders in the 
slow-rotation expansion. In other words, we are interested in a 
simple, practical approach to this kind of problem.
 
As was shown in the previous section, the relativistic r-mode problem 
has (essentially) the following form
\begin{equation}
\Omega y^{\prime\prime\prime\prime} + x y^{\prime\prime} + B y = 0
\end{equation}
in the vicinity of the point $x = r-r_0 =0$  (we use primes to
indicate derivatives with respect to $x$).  Both this toy problem, and
the problem outlined below,
retain the main character of Eq. (\ref{higherorder}) but are sufficiently simple
that we can solve them analytically.
From standard perturbation theory, we know that this class of problems
can be approached via matched asymptotic expansions. Typically, the
outcome is that the singular equation (the equation obtained 
by taking $\Omega\rightarrow 0$) leads to an accurate solution 
well away from $x=0$, while the higher order term is required 
to regularize the solution near the origin. To illustrate this, 
and to motivate the method used to solve the 
r-mode problem in the next section, we consider the 
toy problem  
\begin{equation}
\epsilon y^{\prime\prime\prime\prime} + x^2 y^{\prime\prime}
-x y^\prime + y = 0
\label{toy}\end{equation}
where $\epsilon$ is small in some suitable sense.

Assuming a power series expansion in $\epsilon$ we see that we first need to 
solve the singular equation,
\begin{equation}
x^2 y^{\prime\prime} -x y^\prime + y = 0.
\end{equation}
The two solutions to this equation are $y_1= x$ and $y_2=x \ln |x|$. 
In other words, the solutions to our toy problem are 
similar to the two (local) solutions we found for the relativistic 
r-mode problem in Sect.~II. 
Hence, a method for solving our toy problem should be 
equally valid for the r-mode problem.

\begin{figure}[h]
\centerline{\epsfysize=6cm \epsfbox{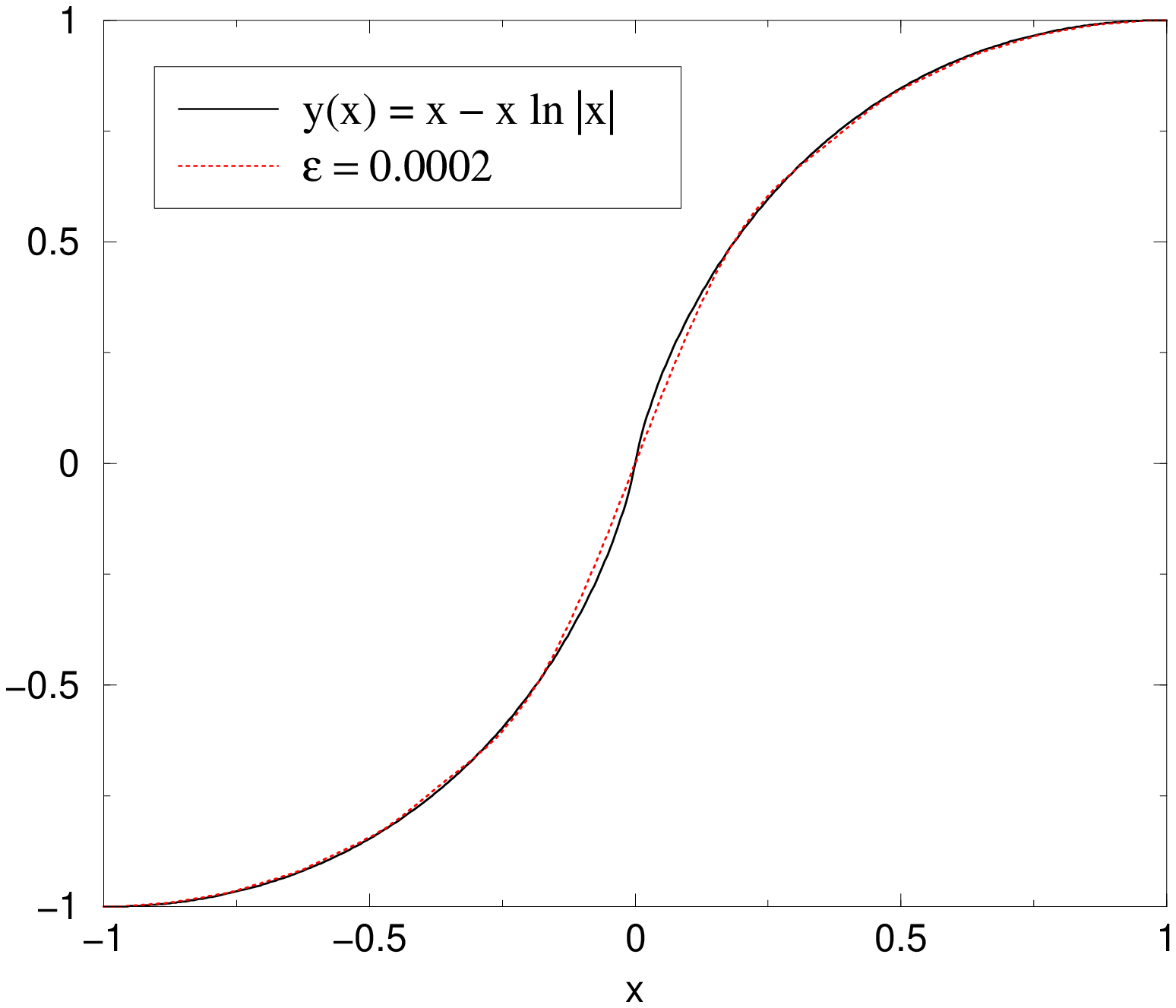} 
\epsfysize=6cm \epsfbox{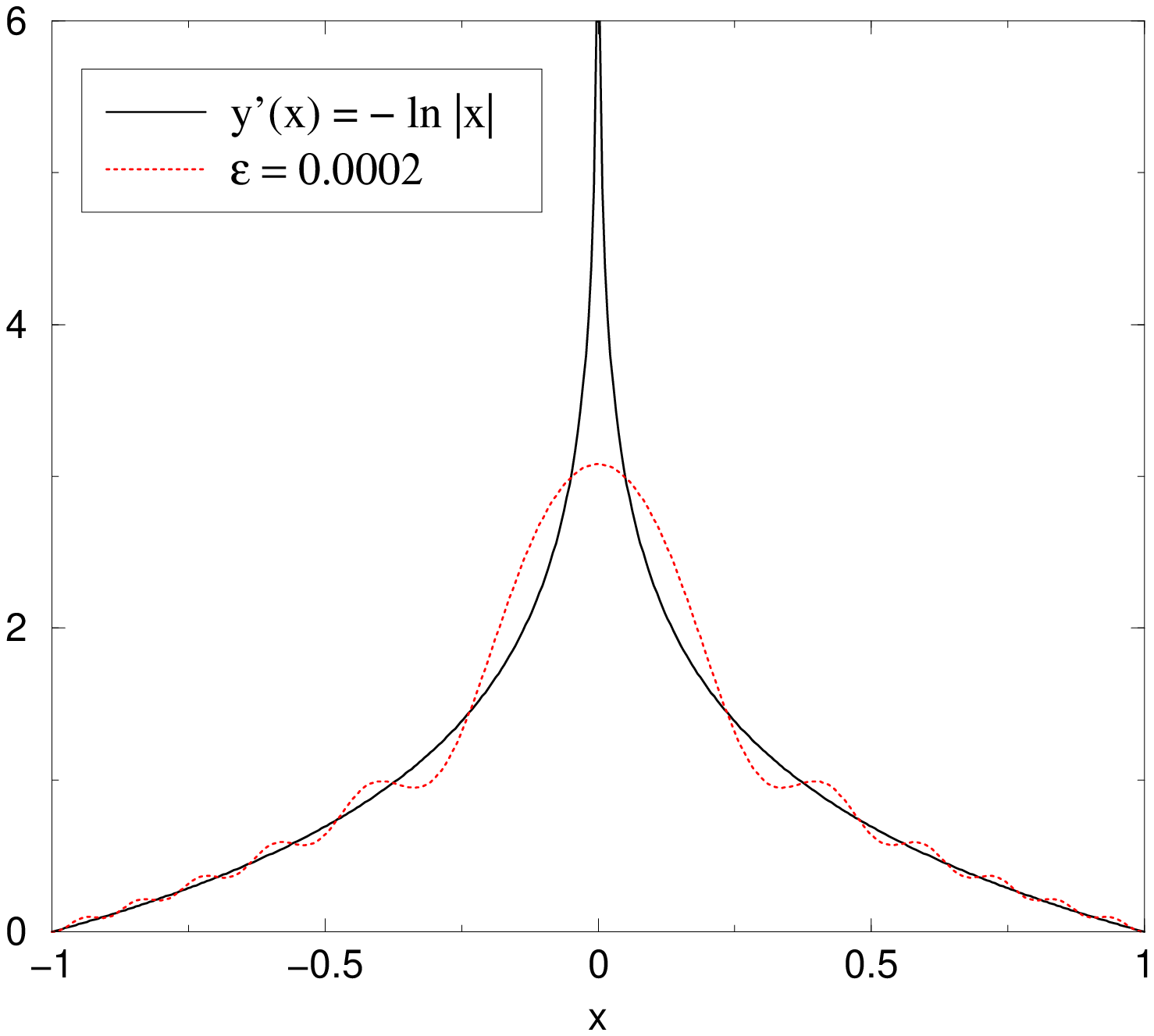}}
\caption{We compare the singular solution of our simple toy problem 
to the complete solution (for $\epsilon=2\times10^{-4}$) of the higher 
order (non-singular) equation. While the two functions agree well 
over the entire range (left panel), the derivatives obviously differ 
near the singular point at the origin (right panel). The figures illustrate 
that the singular solution provides an acceptable approximation to the 
true solution well away from a region near the origin.}
\label{toyfig}
\end{figure}

Let us now suppose that we are interested
in a global solution that satisfies boundary conditions
$y(1)=1$ and $y^\prime(1)=0$. Then we must have 
\begin{equation}
y(x) = x - x \ln|x|
\label{sin}\end{equation} 
As was the case in Sect.~II, this function is well behaved at 
the origin but its derivative diverges (cf. Fig.~\ref{toyfig}).
Note that this solution also satisfies the boundary conditions
$y(-1)=-1$ and $y^\prime(-1)=0$.

Let us now consider the full fourth order equation (\ref{toy}).
It is straightforward to solve it using power series expansions. 
Inserting $y=\sum_{n=0}^\infty a_n x^n $ in Eq. (\ref{toy}) we find the 
recursion relation
\begin{equation}
a_{n+4} = - { (n-1)^2 a_n \over \epsilon (n+4)(n+3)(n+2)(n+1)} 
\end{equation}
From this we see that we have four independent solutions. One of these, 
corresponding to $a_1\neq 0$ truncates and leads to the solution
$y=a_1 x$. 
 As a result of the simple recursion relation, we can 
write the general solution to Eq. (\ref{toy}) in closed form:
\begin{equation}
y(x) = a_0 y_0(x) +  a_1 y_1(x) + a_2 y_2(x) + a_3 y_3(x)
\end{equation}
with
\begin{eqnarray}
y_0(x) &=& \sum_{i=0}^\infty 
\frac{(-1)^i \ [(4i-5)!!!!]^2}{(4i)! \ \epsilon^i} \ x^{4i}
\\
&&\nonumber\\
y_1(x) &=& x 
\\
&&\nonumber\\
y_2(x) &=& x^2 \ \sum_{i=0}^\infty 
\frac{(-1)^i \ 2 \ [(4i-3)!!!!]^2}{(4i+2)! \ \epsilon^i} \ x^{4i} 
\\
&&\nonumber\\
y_3(x) &=& x^3 \ \sum_{i=0}^\infty 
\frac{(-1)^i \ 6 \ [(4i-2)!!!!]^2}{(4i+3)! \ \epsilon^i} \ x^{4i}
\end{eqnarray}
where we have defined the symbol $k!!!! = k (k-4)!!!!$ with 
$k!!!! = 1$ for $k\leq 0$.

We now want to find the specific solution to the fourth order problem 
which satisfies the  
boundary conditions
\begin{eqnarray}
y(-1) &=& -1 \\
y(\phantom{-}1) &=& \phantom{-}1 \\
y'(-1) &=& \phantom{-}0 \\
y'(\phantom{-}1) &=& \phantom{-}0 
\end{eqnarray}
so that it agrees with the second order (singular) solution at the 
boundaries. It is straightforward to show that the required 
solution is $a_0 = a_2 = 0$ and
\bea
a_1 &=& \frac{y'_3(1)}{y'_3(1)-y_3(1)} \\
a_3 &=& \frac{-1}{y'_3(1)-y_3(1)}
\eea
or
\be
y(x) = \frac{x\,y'_3(1)-y_3(x)}{y'_3(1)-y_3(1)} 
\label{exact}
\ee
This solution is compared to the singular solution (\ref{sin}) in 
Fig.~\ref{toyfig}. From the data shown in the figure one can conclude 
that the solution to Eq. (\ref{toy}) is well-described by the singular 
result (\ref{sin}) as long as we stay away from the immediate vicinity 
of $x=0$.

\section{The r-modes of nonbarotropic relativistic stars}

The discussion in the previous two sections 
has crucial implications for our attempt to solve the 
relativistic r-mode problem for nonbarotropic stars. 
Clearly, we can use  our two
solutions to Eq. (\ref{kojima}) to 
approximate the physical solution to the problem away from 
$r=r_0$ even though one of these expansions is technically 
singular at $r_0$. This provides us with the means to 
continue the numerical solution of Eq. (\ref{kojima}) across
$r=r_0$, even though we will not be able to infer the exact 
form of the solution in a thin~\cite{thinlayer} ``boundary layer'' 
near this point. Should we require this information we must carry 
the slow-rotation calculation to higher orders and solve a much more
complicated problem.

We thus propose the following strategy for 
solving the relativistic r-mode problem
in nonbarotropic stars: First integrate the regular solution 
from the origin up to $r=r_0-\delta$, where $\delta$ is suitably small. 
Then use the numerical solution to fix the two constants $a_0$ and $b_0$ 
in the linear combination (cf. (\ref{regsol}) and (\ref{singsol}) )
\begin{equation}
h = h^{\rm reg} + h^{\rm sing}
\end{equation}
Finally, this approximate solution is used to re-initiate 
numerical integration at $r_0+\delta$. 
This approach was first advocated by one of us in a set of circulated but
unpublished notes \cite{nils98}, and the idea was resurrected by 
Ruoff and Kokkotas \cite{kr}.

\begin{figure}[h]
\centerline{\epsfysize=6.1cm \epsfbox{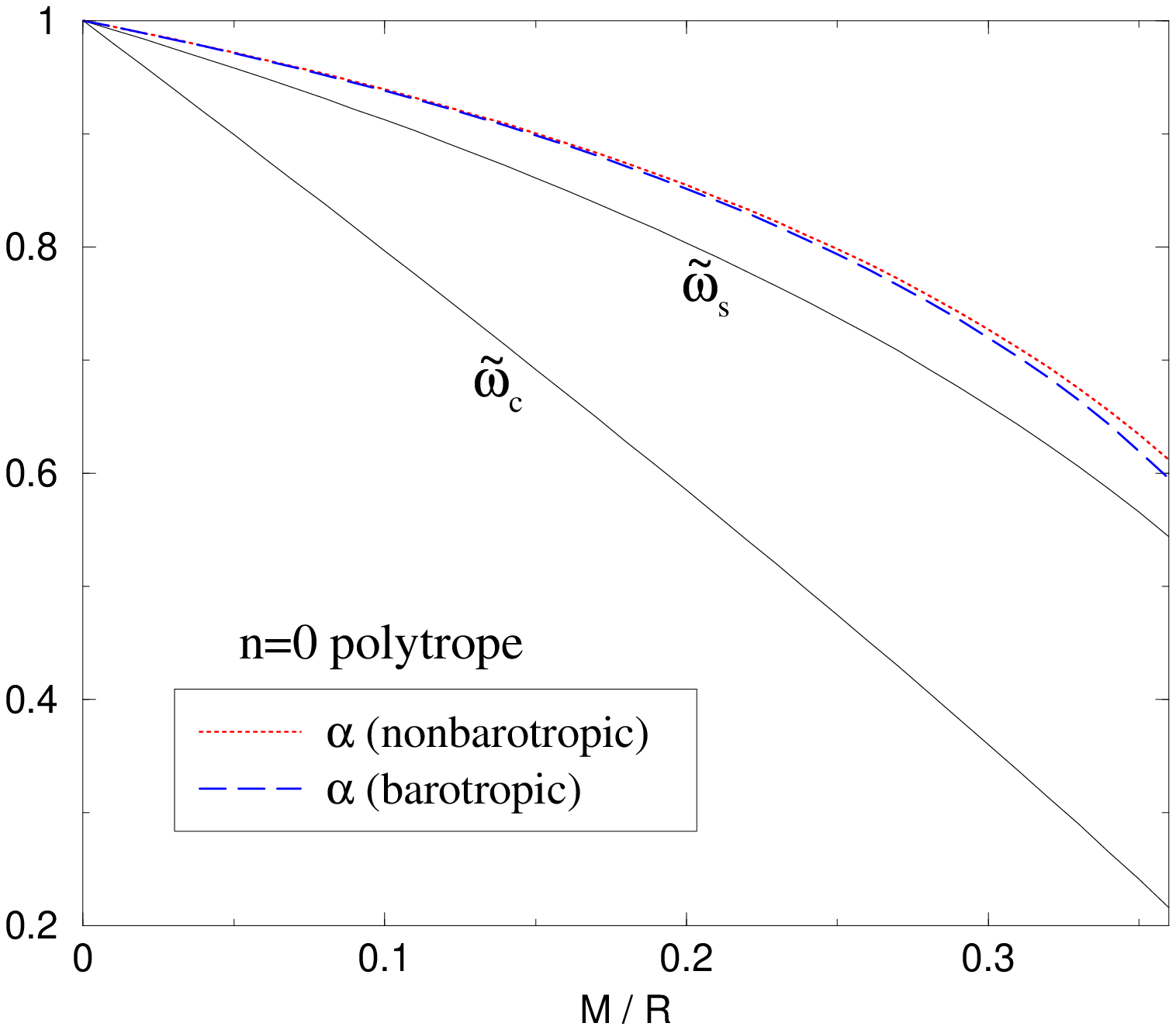} 
\epsfysize=6cm \epsfbox{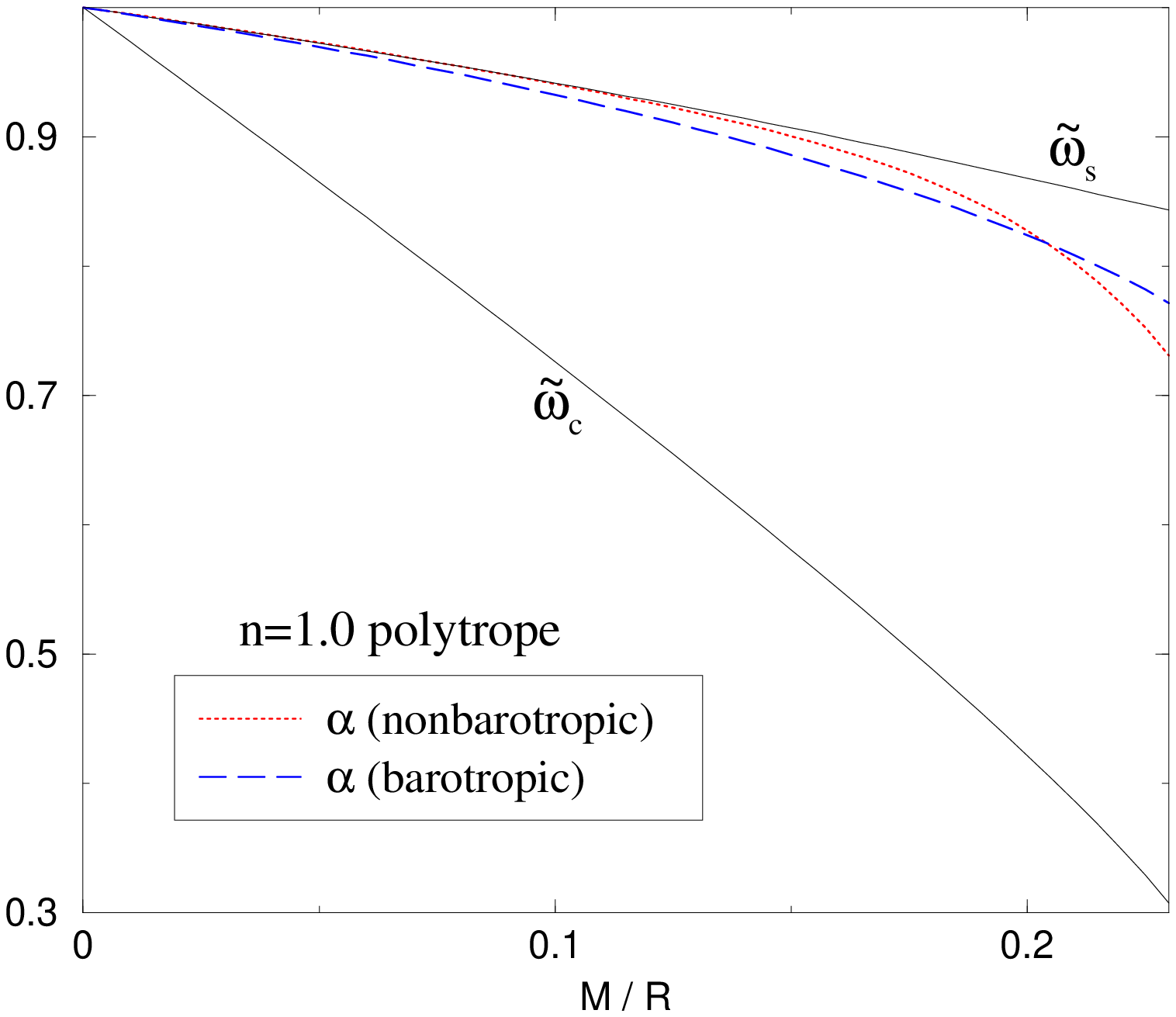}}
\caption{The r-mode eigenfrequencies $\alpha$ of relativistic nonbarotropic 
stars for $n=0$ (left panel) and $n=1$ polytropes (right panel). Also shown 
are the corresponding values of the relativistic framedragging at the centre 
$\tilde{\omega}_c$ and surface $\tilde{\omega}_s$ of the star. Whenever 
$\tilde{\omega}_c < \alpha < \tilde{\omega}_s$ the problem is formally 
singular. As is clear from the data, the uniform density case ($n=0$) is 
always regular while most of our $n=1$ models are in the singular range.
Also shown (as a dashed curve) are the eigenfrequencies for the 
axial-led inertial mode of a barotropic star that most resembles the 
leading Newtonian r-mode.  Note that the inertial mode problem is never
singular.}
\label{relfig}
\end{figure}

We have used the proposed strategy to calculate r-modes for a 
wide range of polytropic stellar models. 
Typical results are shown in Figure~\ref{relfig}. 
(Shown also for comparison are the inertial mode frequencies of fully 
relativistic barotropes~\cite{lfa}. The inertial modes shown are those 
that limit to the $l=m=2$ r-mode of the corresponding Newtonian 
barotropic model.)
By comparing the obtained mode-eigenvalues $\alpha$ to the values for the 
relativistic framedragging at the centre and surface of the star 
($\tilde{\omega}_c$ and $\tilde{\omega}_s$, respectively), one can 
see that the r-mode problem is always regular for uniform density stars. 
As the equation of state becomes softer ($n$ increases) the situation changes.
For example, for $n=1$ polytropes one must typically consider the singular
problem in order to find the relativistic r-mode.
This conclusion is in agreement with Kokkotas and Ruoff \cite{kr} as 
well as Yoshida \cite{yosh}.  It is worth emphasizing that 
the inertial mode problem for
barotropic stars is never singular \cite{laf1,lfa} unless one makes
the Cowling approximation \cite{khp,rsk}, an approximation that is not
in fact appropriate in the barotropic case. 

Before discussing our results further we need to comment on a
difference between our calculation and those in \cite{kr,yosh}. 
In these papers the authors consider polytropic equations of state of the form
\begin{equation}
p = K \epsilon^{1+ 1/n}
\label{krpoly}\end{equation}
with $p$ the pressure and $\epsilon$ the energy density.
Meanwhile, we are using
\begin{equation}
p = K \rho_0^{1+1/n} \quad , \mbox{ and } \quad \epsilon = \rho_0 + np
\end{equation}
where $\rho_0$ is the rest-mass density,
in order to stay in line with the analysis of the inertial
modes of barotropic stars \cite{lfa}. 
This means that our numerical results cannot be directly 
compared to those in \cite{kr}. In order to verify that the results
are consistent we have done some calculations using also (\ref{krpoly}). 
We then find that our results are in perfect agreement with those of 
Ruoff and Kokkotas.

Our calculations thus support the numerical results of the 
previous studies. It is clear that, for more realistic equations of state
one must consider the singular r-mode problem. Where we differ
from both Ruoff and Kokkotas \cite{kr} and Yoshida \cite{yosh} is in the 
interpretation of 
the results in these cases. Yoshida only considers the regular problem, and
tentatively argues that there may not exist any relativistic r-modes 
when the problem is singular. Similar conclusions are drawn by 
Ruoff and Kokkotas \cite{kr}. 
As we have already indicated, we disagree with these 
conclusions.  Even in the slow rotation approximation the
perturbations, obtained by solving the initial value problem, are
non-singular.  However, the root cause of the singular nature of the
mode problem 
is a breakdown in the slow-rotation approximation. 
We believe that this problem would not arise if the calculation were taken to 
higher orders in $\Omega$ in the vicinity of the ``singular'' point
(in analogy with boundary layer studies in problems involving viscous
fluid flows). The {\em physical} problem is likely to be perfectly 
regular, but unless we extend the slow-rotation calculation to 
higher orders
(or approach the problem in a way that avoids the slow-rotation expansion)
we cannot solve the r-mode problem completely for nonbarotropic stars.
However, we have shown how the r-mode eigenfrequencies can be estimated using 
only the solution to the singular mode problem, where they manifest
themselves as zero-step solutions. 

\begin{figure}[h]
\centerline{\epsfysize=6cm \epsfbox{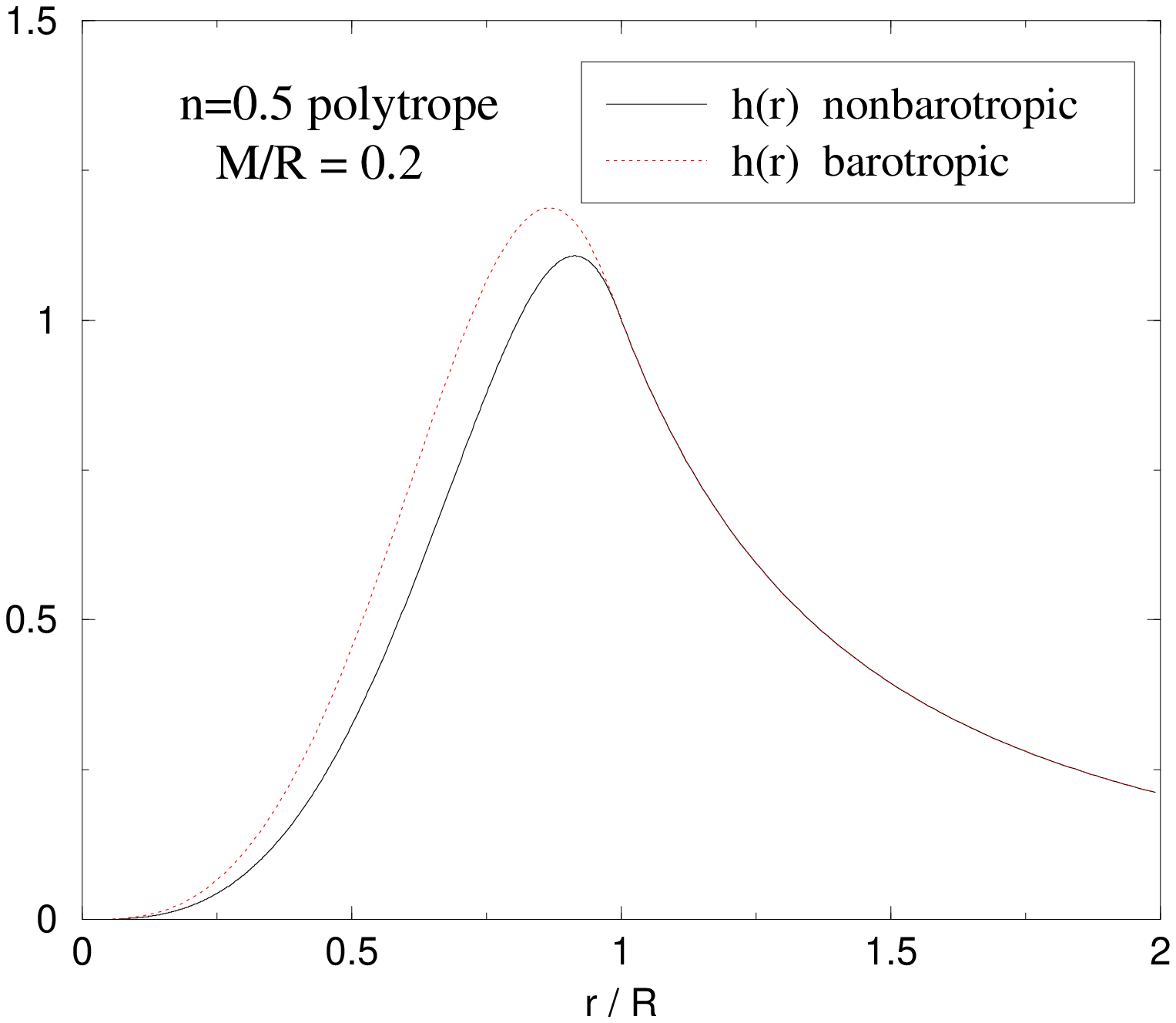} 
\epsfysize=6cm \epsfbox{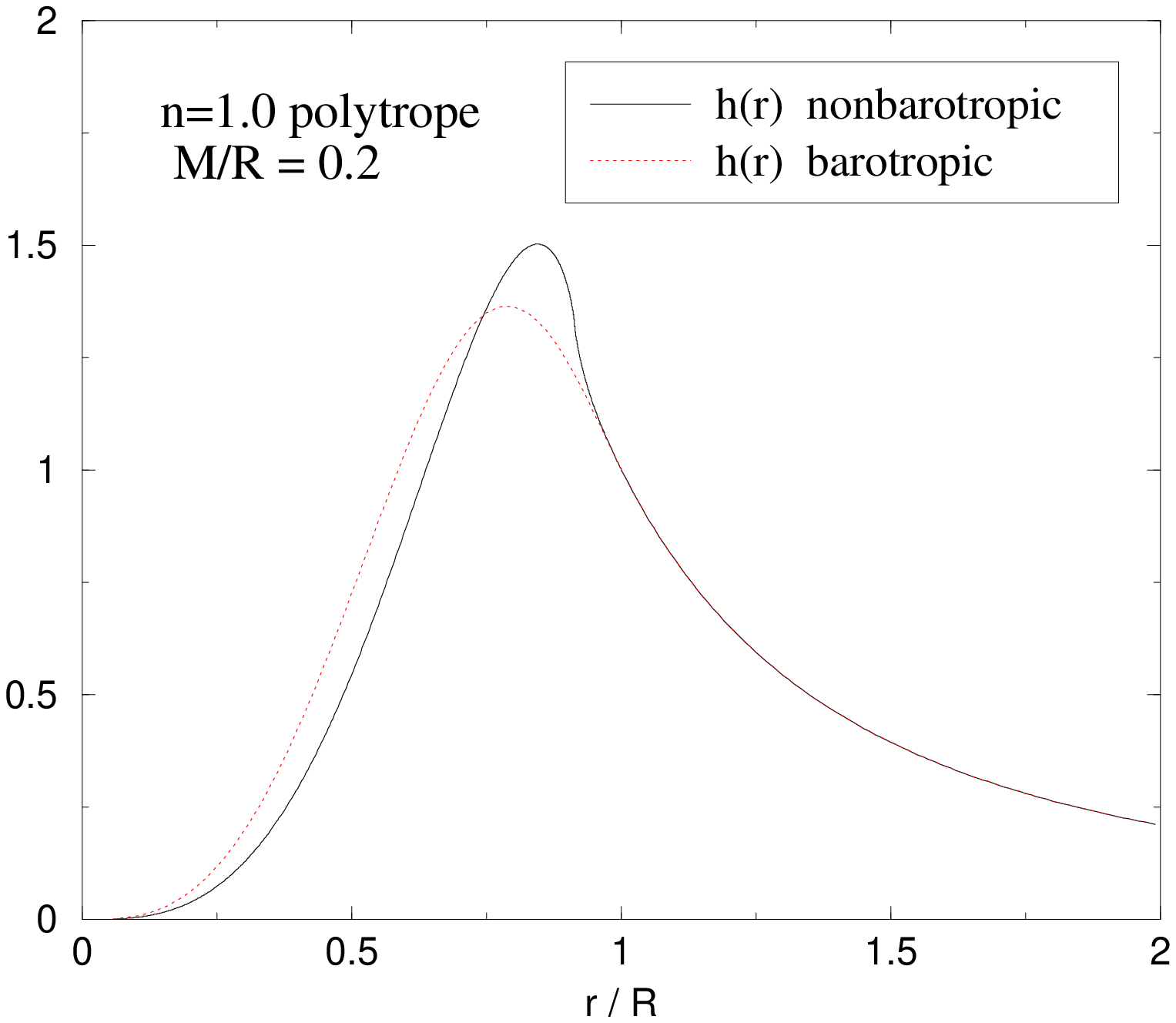}}
\caption{The eigenfunction for an r-mode of a relativistic nonbarotropic
star is compared to the corresponding axial-led hybrid mode of a barotropic 
model. In each comparison, the equilibrium model is chosen to be the same: a
relativistic polytrope of compactness $M/R=0.2$ and polytropic index 
$n=0.5$ (left panel) or $n=1.0$ (right panel). The left panel shows a case 
in which the nonbarotropic mode is regular while the right panel shows
a case in which the nonbarotropic mode is singular (the singular point is 
close to the surface at $r_0 = 0.913 R$.) 
The functions are all normalized so that $h(R)=1$.}
\label{funcfig}
\end{figure}

The case in favor of our approach has been argued (we believe 
convincingly) in the previous sections. In addition, we can provide one further
piece of evidence. In our previous study \cite{laf1}, it was pointed out 
that there is a striking similarity between the eigenfunctions of modes in
barotropic and nonbarotropic stars.  For example, the metric variable
$h(r)$ for an $l=m=2$ r-mode of a nonbarotropic uniform density 
star was very similar to that of the axial-led hybrid mode
corresponding to the Newtonian $l=m=2$ r-mode. This is exactly what one would 
expect if the two represent a related physical mode-solution. We can now 
extend this comparison to the case of polytropic stars. The relevant data are 
shown in Figure~\ref{funcfig}. 
We believe these data provide further support for the relevance of our 
nonbarotropic relativistic r-mode results.

\section{Conclusions and caveats}

We have discussed the calculation of r-modes of relativistic nonbarotropic
stars, shedding new light on a problem that has been associated with some
confusion in the literature. We have shown how the seemingly 
singular problem can (in principle) be regularized, using standard ideas from 
boundary layer theory and viscous fluid flows, and how one can
nonetheless estimate the 
eigenfrequencies of the desired r-modes from the singular mode
problem.  There are however issues that remain to be resolved, two of
which merit particular comment.  

Kojima's equation admits a
continuous spectrum of singular solutions whose collective
perturbation is  non-singular.  The time-dependence of the collective
perturbation is complicated, but includes a position dependent
frequency contribution, and possible power law decay with time.
At certain frequencies within the continuous spectrum one can find
perturbations that behave like stable modes, whose 
manifestation is again non-singular (the zero-step solutions).  We have argued in this paper
that the underlying physical problem can  be regularized by
considering higher order rotational corrections.  The effect of such
regularization on the continuous spectrum and zero-step solutions is
as yet unclear.  If the zero-step solutions become regular normal
modes then this would be physically interesting. The fate of the
rest of the continuous spectrum is unknown; it may remain, vanish, or break up
into discrete normal modes.  

The second issue is of relevance should we want to assess
the astrophysical importance of the r-modes we have computed. In order to 
do this we need to estimate the timescale on which the mode grows due to
gravitational wave emission \cite{lfa,rk2}. This calculation 
requires knowledge of the perturbed fluid velocity in order
for the relevant canonical mode-energy to be evaluated. In the notation of 
\cite{laf1}, we need the variable $U(r)$. We know 
from Eq.~(4.21) in \cite{laf1} that
$$
(\alpha -\tilde{\omega}) \ U = -\alpha h
$$ 
Clearly if we were to use our mode solution to Eq. (\ref{kojima}),
the corresponding result for $U$ would necessarily be singular, blowing up like $1/(\alpha -\tilde{\omega})$ at the singular
point. In accordance with the arguments in Sections~II and III above, 
it is easy to argue that the ``physical'' solution $U(r)$ will be
smoothed out by including higher order terms near the singular point and
thus be regular at all points inside the star. However, solving this
higher order problem is difficult.  We can in principle avoid having
to solve the higher order problem by solving instead the
time-dependent initial value problem for the physical velocity
perturbation.  The physical velocity perturbation, just like the
metric perturbation, will be non-singular.  Unfortunately solution of
the initial value problem is very difficult if one does not have a
full analytic solution for the singular mode problem (see \cite{w2}
where the same issues are discussed for the differential rotation problem).  
This may well mean that we cannot meaningfully estimate 
the gravitational radiation reaction timescale for the singular
nonbarotropic modes discussed in this paper.

\acknowledgments

KHL acknowledges with thanks the support provided by
 a Fortner Research
Fellowship at the University of Illinois and by the Eberly research funds
of the Pennsylvania State University. This research was also supported
in part by NSF grant AST00-96399 at Illinois and NSF grant PHY00-90091
at Pennsylvania.
NA is a Philip Leverhulme Prize fellow and also acknowledges support 
from the 
EU programme ``Improving the Human Research Potential and the Socio-Economic
Knowledge Base'' (research training network contract HPRN-CT-2000-00137).
ALW is a National Research Council Resident
Research Associate at NASA Goddard Space Flight Center.

\appendix*
\section{Continuous spectra in the barotropic problem}

Although in this paper we focus on the nonbarotropic problem, a brief
note on the barotropic problem is in order.  The barotropic
normal mode
problem is not singular at lowest order in the slow rotation expansion
if one includes the metric perturbations \cite{laf1,lfa}.  It is however singular
if one makes the Cowling approximation
\cite{khp,rsk}. The Cowling approximation problem can be regularized by considering the
initial value problem \cite{khp} or by coupling to higher order
multipoles \cite{rsk}.  However, Lockitch, Andersson and Friedman
\cite{laf1} have
shown that the Cowling approximation is not appropriate to describe
the inertial modes of a barotropic star. The singular problem that
arises in this case is thus an
artifact associated with an unphysical assumption.  

In Ref.~\cite{rsk}, Ruoff, Stavridis and Kokkotas study
barotropic inertial modes in the Cowling approximation. They expand
the eigenfunctions in terms of spherical harmonics, which leads to a
set of coupled equations that they truncate at some value $l_{max}$ of
the angular parameter $l$.  For a
given $l_{max}$, they find certain frequency bands for which
the matrix problem cannot be inverted, and claim (correctly) that these 
frequency bands represent continuous spectra.  When $l_{max}$ is increased,
these continuous spectra are replaced by a discrete eigenfrequency
solution at a frequency close to that previously occupied by the
continuous spectrum.  However, other continuous spectra now appear at
different frequency bands.  As $l_{max}$ increases these continuous
spectrum bands grow in number and begin to span the full range of 
frequency space. 

The authors do not explain why this should be
the case, but note that the continuous spectra may
vanish when further higher order couplings are taken into account.  In
fact they should also vanish in the limit $l_{max}\rightarrow \infty$
with only the lower order couplings that they consider.  The
continuous spectra that they observe for a  
given $l_{max}$ are the continuous spectra associated with the
inertial modes of highest $l$.  By including one more term in the
coupling equations these solutions are regularized by coupling to
higher order multipoles, hence the appearance of discrete mode
frequencies.  At the same time a new set of continuous spectra appear,
associated with the unregularized inertial modes that have
$l=l_{max}$. As $l$ increases there are more and more inertial modes \cite{lf},
hence the apparent proliferation of continuous spectra.  In
the limit $l_{max}\rightarrow \infty$ the inertial mode 
problem will be regular,and concerns that the frequency band may fill up with
continuous spectra are unfounded even in the low coupling
approximation. In this limit mode "disappearance" within continuous
spectra is no longer an issue.  In the limit of finite $l_{max}$ modes
may however appear as zero-step solutions within the continuous
spectra: this is a topic for further study.


\begin{thebibliography}{10}
\bibitem{orders} It should be noted that in a Newtonian nonbarotropic star 
the r-mode eigenfunctions are fully computable only at 
order $\Omega^2$. In contrast, as 
discussed in this paper, the r-modes of a nonbarotropic relativistic star are
partially computable already at order $\Omega$.
\bibitem{modeparity} If a mode varies continuously along a sequence of 
equilibrium configurations that starts with a spherical star and continues 
along a path of increasing rotation, it is natural to call the mode axial 
if it is axial for the spherical star. Its parity cannot change along the 
sequence, but $l$ is well-defined only for modes of the spherical 
configuration.
\bibitem{thinlayer} The thickness of this 
boundary layer depends on the exact nature of the higher order
terms in the slow rotation expansion, only one of which is 
included in Eq. (\ref{higherorder}).
\end{thebibliography}
\end{document}